# On The Recovery Performance of Single- and Multipath OLSR in Wireless Multi-Hop Networks

Inès Doghri*, Laurent Reynaud*, and Isabelle Guerin-Lassous**

*Orange Labs, Lannion, France
**Lyon University – LIP, Lyon, France
{ines.doghri, laurent.reynaud}@orange-ftgroup.com and
isabelle.guerin-lassous@ens-lyon.fr

**Abstract.** In this paper, we study and improve the recovery properties of single- and multipath routing strategies when facing network failure situations. In particular, we focus our study on two MANET routing protocols: OLSR and its multipath extension MP-OLSR. In various wireless multi-hop network environments, especially in multiple chain topologies, we define and seek to evaluate the latency introduced by these protocols to find a new path after a link failure. Theoretical estimations and simulation results show that, under dual chain-topologies, this latency can be too long and incompatible with the needs of loss and delay constrained applications. As the source nodes cannot detect link failures immediately because of the delay incurred by the well-known nature of link state protocols in general, and of OLSR Topology Control (TC) messages in particular, these nodes keep sending packets along broken paths. We thus study the inconsistencies between the actual network topology and the nodes' own representation. After analyzing the consequences of this long latency, we seek to alleviate these problems with the introduction of adapted mechanisms. We propose three new different schemes and accordingly extend the original OLSR and MP-OLSR protocols in order to decrease the expected latency and improve the protocol performance. Simulation results show a steep decrease of the latency when using these new schemes in dual chain-topologies. We also discuss these results in terms of packet loss, end-to-end delay and overhead.

**Keywords:** ad hoc networks, multipath routing, link state protocols, fault tolerance.

## 1 Introduction

A Mobile Ad Hoc Network (MANET) is formed by a collection of wireless mobile nodes which can dynamically exchange and relay data. Provided there is at least one path from a source to a destination, this pair can thus communicate, even if not in direct radio range. However, while the multi-hop nature of MANETs facilitates the connection between nodes, those networks face many traffic impediments, often provoked by rapidly changing topologies. Some causes for those continuous changes are unpredictable node mobility patterns and fluctuating radio link quality. Moreover, some of those nodes may be energy and bandwidth constrained. For these reasons, paths between source-destination pairs are generally unstable over time and one primary objective of MANET routing protocols is to alleviate such instability by performing optimal path discovery and maintenance. For these protocols, performing an efficient recovery, i.e. limiting data loss and incurred delays after a link failure, is a desirable feature and a challenging issue. In this context, introducing some form of fault tolerance is one of the expected benefits of multipath routing, along with load balancing and route aggregation. Multipath protocols provide backup paths in case unacceptable degradations or failures are detected on the active paths. As a result, faster and more efficient recoveries are meant to follow route failures. However, this efficiency greatly depends on the failure detection speed and on how fast corrective measures are applied by the multipath routing protocol. As an illustration, the authors of [16] experimented on a real MANET made of 18 OLSR nodes. In particular, they introduced or removed nodes and measured the consecutive recovery times by the surrounding nodes. According to their results, it takes about 4 to 10 second for the updated topology information corresponding to an appearing node to be disseminated through the whole MANET. Likewise, a disappearing node event and the subsequent lost link information dissemination take about 7 to 11 seconds to reach the most remote nodes of the considered MANET. Such measurements highlight that a high rate of topological events such as those previously mentioned can result in significant dissemination times and inconsistencies in network views, which can negatively affect the network performance. This study only concerns the single path routing OLSR in the specific context of nodes appearance/disappearance.

In our work, we aim at deepening the study of recovery time and propose new schemes to single- and multipath OLSR routing which are designed to improve the recovery properties as well as the performance of the routing protocols. More precisely, we are interested in their performance in various topologies, among which Wireless Mesh Network (WMN) topologies, where node mobility, compared to that of general MANET topologies, is significantly reduced or null. This context is of utmost importance for operators as an efficient WMN deployment is dependent of network topology, among other important key factors [5]. When deploying WMN, network operators not only seek to create uniform wireless access coverage by judiciously placing WMN nodes on the designated area, they also need to provide adequate backhaul connectivity. For this latter issue, WMN are increasingly seen as an efficient alternative to known wired and wireless backhaul technologies [6], especially when cost-efficiency is sought, for instance in emerging regions, where operators do not necessarily deploy wired backhaul infrastructures. But for a wide acceptance of WMN as backhaul network, it is important that the recovery process does not imply unacceptable delays for some networking applications, like for instance, Telephony over IP (ToIP) or streaming.

The rest of the paper is organized as follows. We first present works related to recovery mechanisms in Section 2. In Section 3, we provide an analysis of the delay recovery related to long latencies produced by OLSR and MP-OLSR in dual chain-topology networks. Then, we propose in Section 4 three different adapted schemes. Afterwards, we discuss performance evaluation results in Section 5. We finally present the concluding remarks and the ongoing work.

## 2 Related Works

Recovery mechanisms largely depend on the nature, generally either proactive or reactive, of single- and multipath protocols. While in proactive protocols routing information towards all nodes are continuously kept up-to-date, reactive protocols build and maintain their routes on demand. However, some protocols, sometimes classified as hybrid, use both proactive and reactive mechanisms: in this case, recovery mechanisms may rely on either or both approaches. In this section, we give a brief description of works related to the proactive, reactive and hybrid recovery mechanisms in case of route failures.

As a general rule, when a proactive MANET protocol experiences a link failure, remote nodes (i.e. nodes that cannot directly sense the failure on one of their communication interfaces) will be unaware of this link loss until they receive a control message carrying the relevant information. The Optimized Link State Routing protocol (OLSR) [1], [4] is an example of link state MANET proactive unipath protocol, for which any link failure will temporarily trigger a topology inconsistency on each remote node, as the topology information of the network is obtained by a periodic dissemination of HELLO and Topology Control (TC) messages. One of the uses of locally exchanging HELLO messages, containing the list of neighbors known by a node and their related link status, is for each node to maintain symmetric links and to build a representation of its one-hop and two-hop neighbors. On the basis of this information, each node independently selects its own set of multipoint relays (MPR) [1] among its one-hop neighbors in order to be able to cover all two-hop neighbors. This neighborhood information has an associated validity time, NEIGHB_HOLD_TIME. As for TC messages, they are disseminated in the whole network by nodes selected as MPR to reduce the number of retransmissions, and contain the address and the sequence number of MPR selectors. From these TC messages, each node of the network builds and updates its view of the network topology. In order to decrease overhead, control messages are not emitted each time topology changes, but rather on a regular basis: time intervals and their default values are defined in [1]. Important intervals, which are used in the rest of this paper, are respectively HELLO_INTERVAL and TC_INTERVAL (the durations between two consecutive emissions of HELLO and TC messages) and MAXJITTER (a maximum additional anti-collision delay added when disseminating control messages such as TC messages).

As a result, not only those messages convey an incurred delay, but also, in the meantime, the topology is susceptible to change because of the dynamic nature of most multi-hop networks, Wireless Mesh Networks (WMN) included. Undesirable factors such as collisions and traffic congestion will further increase such delays . Other induced delays may be intentional (e.g. the OLSR Jitter insertion [1], which is basically an anti-collision mechanism, where the jitter is a value randomly selected in the interval [0,MAXJITTER]), with the same results. Consequently, the design of proactive protocols requires a careful trade-off between the freshness of the link state information on each node of the network and the total overhead induced by the control messages required to maintain this information. This trade-off is more favorable with OLSR, as it uses the concept of MPRs which disseminate control messages with a reduced overhead, compared to classical flooding schemes. Instead of trying to decrease the OLSR control message emission frequency, the authors of [12] devised a Fast TC strategy, according to which TC messages are immediately disseminated after a new topology event. However, a minimum interval between consecutive TC messages is still applied, in order not to significantly worsen the aforementioned trade-off.

Recovery issues are of a different nature with reactive protocols. They generally rely on control messages named Route Error messages [13], [14] and are part of a dedicated maintenance phase. These messages are not disseminated through the entire network, but are precisely sent to the nodes that need to be aware of the changes of topology (e.g. upstream intermediate nodes along a broken route, or relevant source nodes). Compared to the dissemination of topology changes performed by proactive routing protocols, reactive recovery is faster and generates less overhead. However, this maintenance phase must be followed by a re-discovery phase in order to rebuild the broken routes, which naturally reduces the overall efficiency of reactive recovery schemes. This default route re-discovery mechanism is likely to induce overhead and longer packet delays. To reduce this additional overhead, a fast local recovery scheme named Proximity Approach To Connection Healing (PATCH) was introduced in [17]. It relies on the emission of a reactive request message and initiates a new end-to-end route discovery only if there is no repair route found by the node before the link failure. When an intermediate node detects a broken link towards the next hop, it saves the data packet in a local buffer and broadcasts a route request (RREQ) within its 2 hops region, containing the further original intermediate nodes, to quickly repair the route. Any node receiving the local recovery request will send back a local recovery reply if it belongs to the node list. On receiving a recovery reply, the node will transmit the data packet and send both the repaired route and broken link information to the source node. If no reply is received, the data packet is dropped and an error message is sent back to the source. The flooding of this message creates additional control packets. A similar approach, named bypass recovery, seeks to quickly detect broken links and to preserve as much of the original route as possible, without requiring a full rediscovery phase. To do so, it also relies on a fast local recovery scheme that establishes a bypass between the intermediate node that detected the broken link and an alternative node that can connect back to a fragment of the downstream route to the destination. [20] and [21] respectively proposed an extension of Dynamic Source Routing (DSR) [14] and Ad Hoc On Demand Distance Vector Routing (AODV) that supports bypass recovery and showed that these schemes, compared to their original extension, exhibit a minimized overhead and a better throughput.

The performance behavior of different route recoveries of AODV, based on the source repair by initiating the route re-establishment, or on the local repair when the node before the link break handles the route recovery process, are observed and compared to the original repair scheme of the AODV protocol Request For Comments [19] which is based on a composite implementation of both schemes in [18]. Depending on the scenario, the simulation results show firstly that the source repair performs better than the local repair for packet delivery fraction and secondly that the local repair induces less overhead and delay than the source repair. Thus, it is preferable to choose a suitable route recovery mechanism according to the network topology and user application than adopting the composite implementation specified in [19].

Recovery mechanisms may also use both proactive and reactive schemes. That is the case of MP-OLSR [2], [3], which inherently uses the same proactive mechanisms as the OLSR protocols it extends, with some added on-demand mechanisms. MP-OLSR, which is a hybrid multipath protocol, has link state properties, because it reuses OLSR to disseminate and build the topology information. However, routing is performed differently: routes are computed when there are data packets to emit and are built at the source node, with the available link state information. The computation of multiple routes uses the Multipath Dijkstra (MP-Dijkstra) algorithm. MP-Dijkstra, which extends the Dijkstra shortest-path algorithm, selects multiple paths according to the information gathered by the topology sensing mechanisms and can be configured to obtain either node- or link-disjoint paths. OLSR packets are extended with a MP-OLSR header containing the list of intermediate nodes to the destination. Those intermediate nodes will forward packets accordingly, after verifying in their link state information that the next hop is reachable. If not, a route recovery phase is initiated. It relies on the multipath nature of MP-OLSR but, if no alternate route is available, packets are eventually dropped. MP-OLSR (and OLSR) recovery performance can be significantly increased by the optional support of OLSR Link Layer Notification (LLN) [1]. In this case, the routing protocol is able to receive notifications from the link layer when a link between a node and one of its neighbors is broken. Such notifications, which are used concurrently to HELLO messages information, significantly decrease the average delay of link loss detection.

In the same way, the authors of [15] explore also the possible routing pathologies and failures of OLSR in congested networks. In this context, they propose a hybrid routing protocol which is a combination of OLSR with a Reactive Route Recovery (OLSR-R3) process. Since the HELLO packets are sent periodically and are not directly emitted after a failure, this can increase route recovery delay. To speed up the recovery process, R3 initiates a new route discovery by broadcasting a RREQ message to its neighbors in the same manner as AODV [13]. This RREQ is iteratively transmitted by successive neighbors until it reaches the destination. After receiving the RREQ, the destination then sends a unicast route reply (RREP) message to the sender along the reverse path. To setup symmetric links information, the on-demand HELLO packets are encapsulated in RREQ and RREP. The reception of RREQ can ensure that there exists at least a uni-directional link between sender and receiver. However, the embedding of HELLO packets implies larger RREQ messages. As they are flooded across the network, they are thus likely to induce excessive routing overhead. It is to be noted that simulations were performed in a static network environment and do not discuss several important metrics such as the induced control overhead and the impact of this hybrid scheme on the network behavior.

## 3  Analysis of long recovery delays in dual-chain topologies

The long recovery times measured in the OLSR testbed proposed in [16] can negatively affect the network performance, as previously mentioned in Introduction.

To explain the several reasons of these inconsistencies, in this section, we want to evaluate this recovery time under other topologies for OLSR as well as for its multipath version. To do so, we intend to investigate the performance of multi-hop protocols with simple network configurations, such as the dual chain-topology presented in Fig. 1. We were particularly interested in multi-hop chains [11], because that kind of topology is often used when deploying a backhaul network, between the access network and the core network. In many practical cases, nodes use directive antennas and only communicate with a very definite set of peer nodes, generally a predecessor and a successor along the backhaul path.

In the rest of this document, we define latency by the delay between the time corresponding to the first data packet dropped due to a specific link failure, and the time when sources (i.e. nodes that emitted the data packets that were dropped as a direct result of this link failure) effectively recomputed their routing table accordingly. However, in this section, we simplify this definition by focusing on a scenario with 8 nodes, only one source-destination pair (node 0 being the source and node 4 being the destination) in a dual chain-topology without mobility, as depicted in Fig. 1. We suppose a specific failure event F occurs on link (2,3) at time $T_f$ and in this context, we formalize latency as:

$$\Delta = |\ T_r - T_d\ |. \tag{1}$$

Where
- $T_d$: the time when a first data packet, initially emitted by node 0, is dropped by node 2 as a result of the failure event F. Note that $T_d$ may be different from $T_f$.
- $T_r$: the time when node 0 has recomputed its routing table, taking into account the topological effects of the link failure event F.

We now want to estimate $\Delta$ when using OLSR as routing protocol in this topology. For simplification purpose and for compatibility issues with the default MP-OLSR scheme, we assume that LLN support is available. According to the topology depicted in Fig.1, we can distinguish two cases:

a) Node 2 sends a TC message taking into account the link loss

Node 3 initially belongs to node 2's MPR Selector Set (i.e. the main addresses of the nodes which have selected node 2 as MPR). After the occurrence of the failure event F, node 2 stops receiving HELLO messages from node 3.

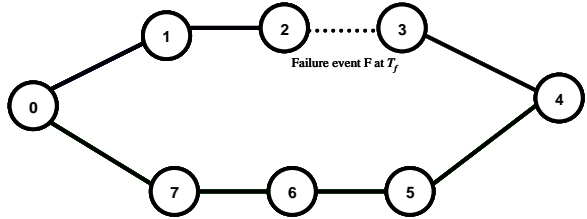

**Fig. 1.** Dual chain-topology: 8 nodes, 2-path scenario, one source/destination pair (node 0, node 4). At time $T_f$, a failure event F occurs on link (2,3).

This latter node will then be removed from node 2's MPR Selector Set after the duration NEIGHB_HOLD_TIME since the last update of node 3's entry. As a result, node 2 cannot send any TC message taking into account this link loss before an elapsed duration of NEIGHB_HOLD_TIME – HELLO_INTERVAL where the default value of NEIGHB_HOLD_TIME is 3 x HELLO_INTERVAL. Moreover, such a TC message will be emitted before a maximum delay which consists of 2 parts: TC_INTERVAL and MAXJITTER. When neglecting the transmission delays, and considering for simplification purpose that a node's routing table is updated at the exact time of reception of a non-empty TC message, we can formalize latency for this first case as follows:

$$\Delta_1 \in [\text{NEIGHB\_HOLD\_TIME} - \text{HELLO\_INTERVAL}, \text{NEIGHB\_HOLD\_TIME} + \text{TC\_INTERVAL} + \text{MAXJITTER}] . \qquad (2)$$

Note that when node 2 detects a failure related to link (2, 3), this link will eventually expire and be removed from the node's Local Link Information Base. This expiration will also trigger a removal of node 3 from node 2's MPR Selector Set, and further emitted TC messages will take into account this change of topology. However, the induced estimated latency, which we will not detail, is included into the delay interval expressed in (2).

Furthermore, the estimation given by (2) is only valid if a TC message carrying the relevant topology information is received by node 0. But several factors such as packet loss may preclude this reception. Another factor is tightly related to the considered chained topology: whenever node 2 emits a TC message that contains the link loss information, node 1 is susceptible to have already been removed from node 2's MPR Selector Set. As a result, the considered TC message is susceptible to be empty. But, according to [1], the emission of such empty TC message is not mandatory. Many OLSR implementations, including those used in the following sections, do not send empty TC messages. If the source node 0 did not previously receive non empty TC messages from node 2 containing the topology change, then it will not receive those empty TC messages either. In this case, (2) does not apply, and another case must be considered.

  b) Expiration of outdated topology information into node 0's Local Topology Set

When the source node 0 has no means to become aware of the link loss, the related information will persist until expiration into node 0's topology representation. This duration essentially depends on the time of reception of the last TC message from node 2. Considering for simplification purpose that node 0's routing table is updated at the exact time of the mentioned information expiration:

$$\Delta_2 \in [2 \times \text{TC\_INTERVAL}, 3 \times \text{TC\_INTERVAL}] . \qquad (3)$$

When using the default values of the parameters defined in [1], (2) and (3) can be expressed as:

$$\Delta_1 \in [4 \text{ s}, 11.5 \text{ s}] \text{ and } \Delta_2 \in [10 \text{ s}, 15 \text{ s}] . \qquad (4), (5)$$

Those estimations show that in this topological context, $\Delta$, whether related to either cases previously discussed, is always significant. On a practical level, it is unacceptable for delay constrained applications. Moreover, for the reasons mentioned in the previous paragraph, both cases expressed by (2) and (3) can occur in this topology.

We verified these estimations with a series of simulations performed on the ns-2 network simulator [7]. These simulations represent the first scenario and are based on the topology described in Fig. 1. The 8 nodes, which are placed into a 150 m x 150 m area, have a transmission range of 60 m. The channel capacity of mobile hosts is set to 11Mb/s. A CBR (Constant Bit Rate) source at node 0 has a 100 kb/s rate with packet size of 512 bytes. The traffic, which is directed at node 4, starts at 10 s and the total simulation time is 50 s. Link failure event F occurs on link (2, 3) at time $T_f$, which varies between 15 s and 19 s. Node 2 being 2 hops away from the source node 0, the topology change (i.e. nodes 2 and 3 stopped being direct neighbors) cannot be known by node 0 either by direct link sense or by HELLO message reception. Thus, node 0 will not be aware of this change until it receives a TC message with the relevant information. In this paper, we used the UM-OLSR [9] and MP-OLSR [2] implementations with LLN enabled. Also, for the MP-OLSR implementation, load balancing is based on round robin and is performed on a maximum of two active paths.

With these parameters, we carry out 5 simulations with different $T_f$ randomly chosen in the interval [15 s, 19 s] in order to compute the minimum, the mean and the maximum values of Δ according to (1). The results presented in Table 1 corroborate (2) and (3), as both OLSR and MP-OLSR exhibit latencies between 4 s and 15 s.

**Table 1.** First scenario: latency simulation results.

| Routing schemes | Latency (s) | | |
|---|---|---|---|
| | Minimum | Mean | Maximum |
| Original OLSR | 4.818 | 10.645 | 14.545 |
| Original MP-OLSR | 11.049 | 13.214 | 14.934 |

## 4 Design of efficient recovery schemes

After having identified this long latency in the previous section, we seek to decrease it with the introduction of adapted mechanisms. To do so, we propose three different solutions and we extend accordingly the original OLSR and MP-OLSR protocols. We give here the detailed description of the different proposed schemes.

1) A Route Error (RE) notification strategy: This well-known mechanism, often used in reactive MANET routing protocols, consists in sending a unicast control message (named RERR_NOTIF) to notify the source about the failure. When an active node detects a link breakage thanks to its local link sense mechanism, it generates a RERR_NOTIF message which contains the interface address of the link failure. It sends this unicast message to the source of the data traffic. When receiving a RERR_NOTIF message, a node updates its topology table by removing information related to the corresponding link. This will trigger the computation of a new routing table based on the current topology information base.

2) A Fast TC (FTC) message strategy: upon detection of a failure on one of its interfaces, a node immediately emits a TC message, named FAST_TC, without waiting for the regular TC message interval of emission. In order to reduce the total induced overhead, we define an interval (FAST_TC_INTERVAL) used for further emission of TC messages related to the same failure event. This strategy is similar to the scheme devised in [12]. However, we also bypass the varying delay (i.e. Jitter) between the generation and the emission of a FAST_TC message by the node which detected the failure. As a result, we intend to obtain a faster global topology update. Note that a standard jitter delay is used according to [1] when intermediate nodes forward this control message, as it is necessary to avoid the potential collisions induced by synchronized broadcasts. We set FAST_TC_INTERVAL to 0.5 s by default.

3) Data Re-emission (DR) strategy: We specifically designed this scheme for the multipath MP-OLSR protocol. It is based on the re-emission of a data packet that was dropped by the node which detected the link failure. This re-emission takes place if an alternate route to the destination was not found during the MP-OLSR standard route recovery. In that case, the strategy takes benefit of the MP-OLSR source routing scheme in order to determine a reverse path towards the source. The packet is then sent towards the destination on a route that includes this reverse path. Before forwarding this packet to the intended destination, the source node will update its Local Topology Set with the topology information embedded into the packet and will recompute, if required, its routing table.

Compared to a OLSR-R3 solution [15] that addresses this issue, our designed recovery strategies is intended to introduce fewer control overhead since RE scheme emits only one unicast control message in one direction while DR scheme does not imply extra control packets.

## 5  Simulation and performance analysis

In this section, we propose to study the behavior of the three recovery schemes we designed. As explained in the previous section, OLSR is extended with two schemes (RE and FTC), while MP-OLSR is extended with all the proposed schemes, including DR. Hence, 3 and 4 schemes were respectively evaluated for OLSR and MP-OLSR, default protocols included. Two distinct scenarios (named scenarios 2 and 3) were designed in order to carry out the performance evaluations. Those were performed under the same ns-2 simulation environment and the same OLSR default implementation as seen in scenario 1.

Depending on the scenarios, 4 metrics were used in order to evaluate the performance of the different strategies:
- *Latency*: defined and calculated according to (1).
- *Packet loss rate*: the percentage of the number of dropped packets among the total number of generated packets.
- *Routing load*: evaluates the overhead due to the emission of routing messages within the network in the presence of applicative traffic. $N_{cm}$ being the total number of emitted and transferred control messages; it corresponds to the percentage of $N_{cm}$ to the sum of $N_{cm}$ plus the total number of received data packets.
- *Average end-to-end delay*: the average delay taken by data packets sent from the sources and received by the destinations. Note that it takes into account queuing and propagation delays.

*Scenario 2* extends scenario 1. It is also based on a dual chain-topology, here with 20 nodes: two nodes (a source-destination pair) are interconnected through 2 node-disjoint paths, each composed of 9 intermediate nodes. The parameters are similar to those of scenario 1 in terms of physical, MAC and applicative configuration. In particular, one CBR flow is also associated to the source-destination pair, with the same parameters. However, the difference with scenario 1 is related to the number of hops, $2 \leq n \leq 8$, between the source node and the first intermediate node which senses the link failure on one of its interfaces. Note that in our simulations, broken links are simulated as the consequence of nodes failure. As a practical consequence, and because node 10 is the destination of the data traffic, the case of *n*=9 is not considered so as to avoid an unneeded failure of the destination. For each location of the failure on the route, 6 simulations are carried out, the failure event time $T_f$ taking each discrete value between 20 s and 25 s. The simulation duration is 100 s.

As shown by Figs. 2 and 3, the average latency induced by a link failure occurring *n* hops away from the source node, for default OLSR and MP-OLSR schemes, is consistent with the values already seen in scenario 1, for *n*=2. Here, for all the considered values of *n*, the source node can only be aware of this failure by either reception of TC messages or by expiration of the relevant information into its Local Topology Set. In this case, the latency is contained between 10 and 14 seconds for OLSR, and between 13 and 14 seconds for MP-OLSR. However, the 3 other schemes, which offer faster topology information update, greatly improve latencies. OLSR_FTC enables latencies below 1 s when $n \leq 4$. With $n > 4$, latency increases with *n* and reaches 3 s when $n = 8$ because of the greater number of hops. MP-OLSR_FTC latency never reaches 2 s.

The two other strategies RE and DR, which are based on unicast transmissions, induce the lowest latencies, which do not exceed 1 s with the RE scheme for OLSR. For MP-OLSR, as illustrated by Fig. 3, RE and DR schemes were both implemented and, in either case, the measured latencies are consistently 95 % less than that of default MP-OLSR.

Figures 4 and 5 show that our route recovery strategies, including RE, FTC and DR, significantly decrease the packet loss ratio by about 90% compared to that of both original routing protocols. This result is explained by the nature of the three considered fast repair schemes: the source node is likely to be informed about a link failure soon after the first packet is dropped and will then quickly update its topological information. In contrast, for both default schemes, the source node will continue to try to send data packets through the failed path until it becomes aware of the broken link. However, as previously mentioned, MP-OLSR uses in these simulations a round robin load balancing and two active paths. In case one path fails, this scheme can thus take benefit of another active path so as to send half the traffic to the destination without interruption. This is confirmed by the results illustrated by

Fig. 4 and Fig.5, where default MP-OLSR shows a 30% to 50% decrease in term of packet loss rate, compared to default OLSR. This result is greatly improved by our three MP-OLSR extensions. Most notably, the DR recovery scheme exhibits the lowest packet loss ratio thanks to the data re-emission process which, at the same time, avoids the drop of data packets and informs the source about broken link information.

*Scenario 3* is related to a more general topology with 50 nodes moving according a Random Way Point (RWP) mobility pattern in a 1000 m × 1000 m area. 10 CBR sources emit 10 packets per second, of 512 bytes each. The wireless nodes have a transmission range of 250 meters and the simulation time is 200 s. The remaining physical and MAC configuration is similar to that of previous scenarios. We run 50 simulations with different mobility scenarios.

This scenario is intended to verify that the schemes extending base OLSR and MP-OLSR, which proved interesting in a dual chained-topology as seen with scenario 2, do not significantly worsen the performance of these protocols in a more general topology with mobile nodes, whose speed varies between 1 m/s and 10 m/s. In particular, as the route error notification and fast TC message strategies are designed to generate additional control messages, we sought to evaluate this extra overhead. Fig. 6, which relates to MP-OLSR, confirms that a limited additional routing load is induced by RE scheme. However, DR repair mechanism reduces the routing load about 2-3% of the original MP-OLSR at mobility speed cases beyond 5 m/s, as it is not based on the emission of additional control messages. The routing load differential slightly increases with node speed.

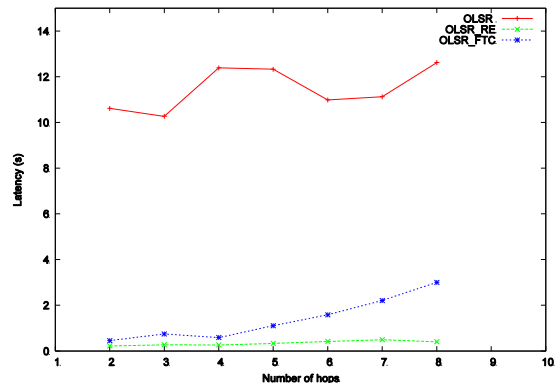

**Fig. 2.** Average latency with the number of hops to the node that senses the link failure, for OLSR, OLSR_RE and OLSR_FTC in scenario 2.

Naturally, MP-OLSR_FTC shows the largest increase, because of the flooding nature of TC messages, unlike route error messages, which are based on unicast transmissions. MP-OLSR_RE thus shows small overhead, which, when combining the results of scenario 2, confirms that the FTC scheme is less efficient in all the studied topologies than the RE and DR schemes. Results, which are not presented here, are similar for OLSR, with smaller routing load differences between the schemes: as OLSR is a single path routing protocol, less control messages need to be emitted for a given number of applicative flows, compared to MP-OLSR.

Fig. 7 shows the average end-to-end delay. The multipath routing, in general, reduces the queue delay because the traffic is distributed along different paths. On the other hand, it might increase the propagation delay because some of the packets are sent through longer alternate routes. When the mobility speed of nodes is less than 4 m/s, the various versions of MP-OLSR, default scheme included, display a similar performance. As again illustrated by Fig. 7, for speeds greater than 4 m/s, the DR strategy tends to display slightly higher end-to-end delays than the other MP-OLSR extensions. In these scenarios, the context of unacknowledged applicative traffic (CBR data over UDP) especially highlights the existence of a trade-off between the loss rate and the end-to-end delay. However, these end-to-end delay results would differ if data retransmissions were required (e.g. traffic over TCP). In this case, the default MP-OLSR scheme, exhibiting the greatest ratio of dropped packets, would be more heavily penalized than its extensions, and especially than the DR extension, which exhibits the lowest packet loss rate.

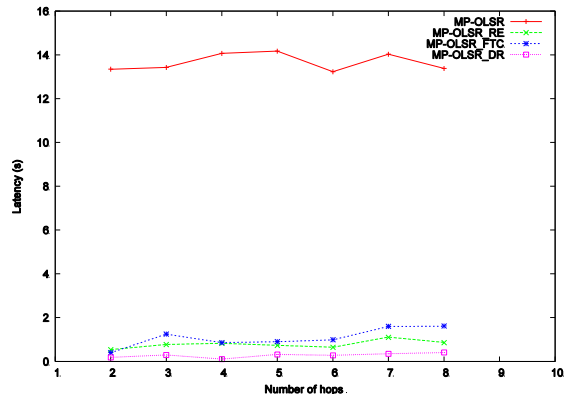

**Fig. 3.** Average latency with the number of hops to the node that senses the link failure, for MP-OLSR, MP-OLSR_RE, MP-OLSR_FTC and MP-OLSR_DR in scenario 2.

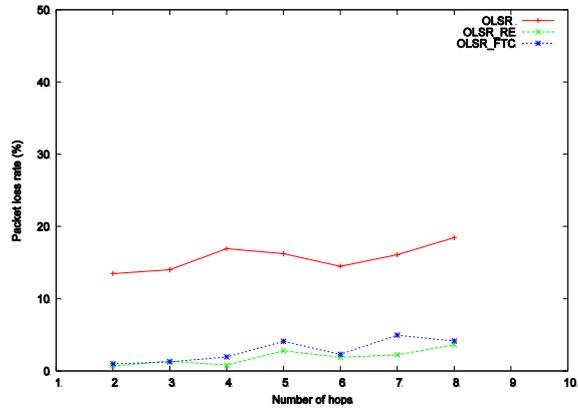

**Fig. 4.** Packet loss rate with the number of hops to the node that senses the link failure, for OLSR, OLSR_RE and OLSR_FTC in scenario 2.

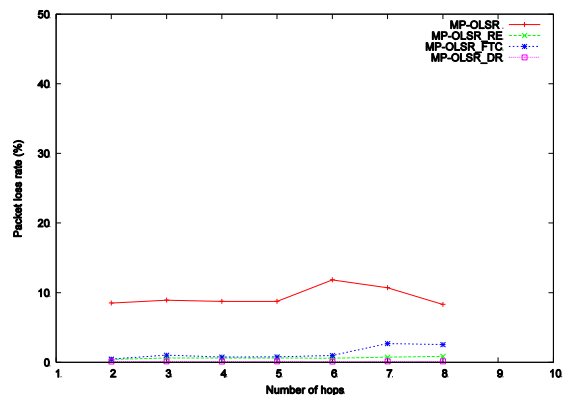

**Fig. 5.** Packet loss rate with the number of hops to the node that senses the link failure, for MP-OLSR, MP-OLSR_RE , MP-OLSR_FTC and MP-OLSR_DR in scenario 2.

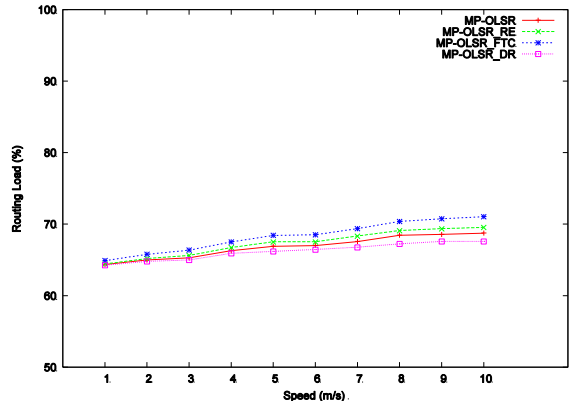

**Fig. 6.** Routing load with node speed, for MP-OLSR, MP-OLSR_RE, MP-OLSR_FTC and MP-OLSR_DR in scenario 3.

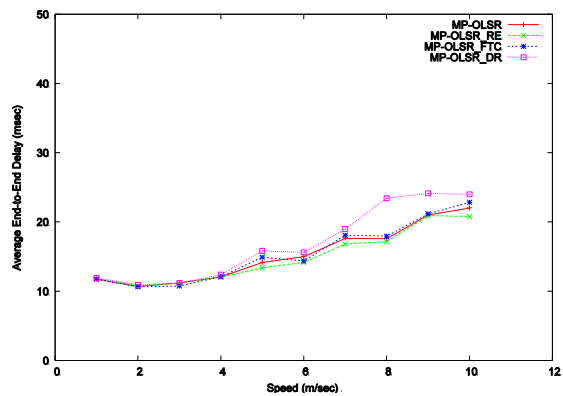

**Fig. 7.** Average end-to-end delay with node speed, for MP-OLSR, MP-OLSR_RE, MP-OLSR_FTC and MP-OLSR_DR in scenario 3.

As the nodes mobility increases, MANET will suffer more from link errors. This is illustrated by Fig. 8, where the packet loss ratio increases with nodes speed. At a speed of 1m/s, the performance of the 3 MP-OLSR schemes is similar to that of default MP-OLSR. However, when faster mobility speeds are considered, it can be observed that the three extensions display a lower packet loss ratio. When a link breaks, the DR scheme avoids dropping packets in case no repaired route to the destination is available: they are sent towards the destination back via the source, thus improving the packet loss ratio as much as 7%. This enhancement is also explained by the reduction of the potential congestion due to the emission of additional control packets, which is particularly important for loss-intolerant applications.

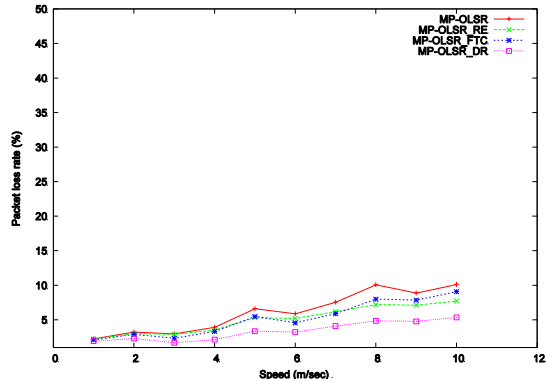

**Fig. 8.** Packet loss rate with node speed, for MP-OLSR, MP-OLSR_RE, MP-OLSR_FTC and MP-OLSR_DR in scenario 3.

## 6 Conclusion and future work

In this paper, we identified a recovery performance issue with link-state protocols in a multi-hop dual chain-topology network. We considered one single path protocol, OLSR, and one of its multipath extensions, MP-OLSR. We studied the different recovery delays consecutive to a link failure and observed that this delay, under several topologies and mobility scenarios, was significant and incompatible with delay constrained applications. We devised 3 adapted schemes to achieve fast recovery: FTC (where a TC message is immediately emitted after a link failure detection), RE (where the node that detected a link failure sends back a unicast message to the source) and DR (where the same node, in the same situation, re-emits the data packet to the source instead), the last scheme being specific to MP-OLSR.

Simulations show that RE is the best scheme for OLSR: recovery performance is substantially improved in a dual chain-topology network, while the overhead induced by the extra control messages is limited in a more general topology, compared to TC. Also, in the same topology, the proposed extensions to MP-OLSR greatly improve the default multipath-based failure repair efficiency in terms of latency and loss rate. For MP-OLSR, DR is the most interesting scheme, as it displays, compared to the two other extensions, a similar or slightly better performance in a dual chain-topology network, while behaving better than the other schemes in terms of both packet loss ratio and routing load in a general topology. Moreover, the combination of default MP-OLSR route recovery mechanisms with data re-emission of the DR scheme enables to avoid dropping data packets and takes benefit of MP-OLSR hybrid link state plus source routing schemes.

In the future, we would like to evaluate those schemes in other realistic topologies and further study delay issues when re-emitting data packets with MP-OLSR and its DR extension. In particular, we intend to study delays and packet-processing mechanisms at a DR source node, when flow-related Quality of Service requirements are known by this source node.